\documentclass[twocolumn]{aa}  	
\usepackage[utf8]{inputenc}
\usepackage{color}
\usepackage{amsmath}
\usepackage{amssymb}
\usepackage{graphicx}
\usepackage{natbib}
\bibpunct{(}{)}{;}{a}{}{,}
\usepackage{hyperref}
\usepackage{txfonts}
\begin{document} 
   \title{Numerical models for the dust in RCW\,120.}

   \titlerunning{RCW~120}
   \authorrunning{A. Rodr\'iguez-Gonz\'alez et al.}
   \author{A. Rodr\'iguez-Gonz\'alez
          \inst{1,} \inst{2},
 Z. Meliani\inst{1},
 M. S\'anchez-Cruces\inst{3},
 P.R. Rivera-Ortiz\inst{2},
   A. Castellanos-Ram\'irez\inst{2}
   }

   \institute{LUTH, Observatoire de Paris, PSL, CNRS, UMPC, Univ Paris Diderot, 5 place Jules Janssen, F-92195 Meudon, France.
              \email{Ary.Rodriguez@obspm.fr}
              \and
             Instituto de Ciencias Nucleares, 
Universidad Nacional Aut\'onoma de M\'exico, 
Ap. 70-543, 04510 D.F., M\'exico
\and
Aix-Marseille Universit\`{e}, CNRS, Laboratoire d'Astrophysique de Marseille, F-13388 Marseille cedex 13, France}

   \date{Received XXXXX; accepted xXXX}

  \abstract
   {The interstellar bubble RCW~120 seen around a type O runaway star is driven by the stellar wind and the ionising radiation emitted by the star. The boundary between the stellar wind and interstellar medium (ISM) is associated with the arc-shaped mid-infrared dust emission around the star within the HII region.
   
}
   { We aim to investigate the arc-shaped bow shock in RCW~120 by means of numerical simulations, including the radiation, dust, HII region, and wind bubble.
   
   }
   {We performed 3D radiation-hydrodynamic simulations including dust using the {\sc guacho} code. Our model includes a detailed treatment of dust grains in the ISM and takes into account the drag forces between dust and gas and the effect of radiation pressure on the gas and dust. The dust is treated as a pressureless gas component. The simulation uses typical properties of RCW~120. We analyse five simulations to deduce the effect of the ionising radiation and dust on both the emission intensity and the shape of the shock.}
   { The interaction of the wind and the ionising radiation from a runaway star with the ISM forms an arc-shaped bow shock where the dust from the ISM accumulates in front of the moving star. Moreover, the dust forms a second small arc-shaped structure within the rarefied region at the back of the star inside the bubble. In order to obtain the decoupling between the gas and the dust, it is necessary to include the radiation-hydrodynamic equations together with the dust and the stellar motion. In this work {all these elements are considered together,} and we {show} that the decoupling between gas and dust obtained in the simulation is in agreement with the morphology of the infrared observations of RCW\,120.}
   {}
%

   \maketitle

\section{Introduction} \label{sec:intro}

The Infrared Astronomical
Satellite (IRAS, \citealt{Neugebauer_etal_1984ApJ...278L...1N}) all-sky survey detected
many extended arc-like structures associated with OB-runaway
stars and paved the way for further studies to  reveal the bow shocks that are present around 
these powerful stellar winds \citep{vanBurenetal1995AJ....110.2914V, Perietal_2012A&A...538A.108P}. Furthermore, the mid-infrared (MIR)  observations from Spitzer-MIPSGAL \citep{Carey_etal_2009PASP..121...76C} of interstellar bubbles along the Galactic plane show that the interiors of  HII regions contain dust, and some of them present 24 $\mu$m dust emission near to the central ionising source \citep{Deharveng_etal_2010A&A...523A...6D,  Kendrew_etal_2012ApJ...755...71K, Simpson_etal_2012MNRAS.424.2442S}. 
The dust grains are expected to evacuate the  interiors of the hot bubbles as a result of the mechanical energy and radiation injected by the massive stars. However, recently, infrared observations show a significant amount of 
dust in the inner part of the bubble, particularly sitting
near to the stellar feedback source. Indeed, high-resolution infrared
observations show that HII bubbles contain a significant
amount of dust in their interiors. \citep{Deharveng_etal_2010A&A...523A...6D, Martins_Pomares_2010A&A...510A..32M, Anderson_etal_2012A&A...542A..10A}.  \cite{Everett_etal_2010ApJ...713..592E} explored the possibility that the evaporation of small
dense cloudlets could resupply the hot gas with a new generation
of dust grains. While this
mechanism can explain the presence of dust inside HII bubbles,
 it is not capable of producing the morphology of the
MIR radiation: an arc-shaped and a peaking close to the ionising
source.
\cite[, see also \citealt{1988ApJ...329L..93V}]{Ochsendorf_Cox_2014A&A...563A..65O} proposed that dust grains are expelled from ionised stars by radiation pressure and \cite{2013ARep...57..573P} performed  1D numerical calculations on the dust in the
the boundary between the stellar wind and the interstellar medium (ISM), and  they concluded that dust deviation by itself is not a viable mechanism to adequately explain the arcs observed in 24 $\mu m$ IR emission in HII regions.
One of the more studied interstellar bubbles with a 24 $\mu$m  arc is RCW~120. The RCW~120 bubble is located at 1.3~kpc  \citep{Zavagno_etal_2007A&A...472..835Z}  at  \textit{l}=348$^{\circ}$.3, \textit{b}=+0$^{\circ}$.5 ($\alpha_{J200}$=17$^{h}$09$^{m}$0$^{s}$, $\delta_{J200}$=-38$^{\circ}$24' \citep{Rodgers_etal_1960MNRAS.121..103R} with a radius of 1.9 pc \citep{Anderson_etal_2015ApJ...800..101A}. 
Its ionising star is CD-38$^{\circ}$11636 or LSS 3959, an O8 type star \citep{Georgelin_Georgelin_1970A&AS....3....1G, Avedisova_Kondratenko_1984NInfo..56...59A, Russeil_2003A&A...397..133R, Zavagno_etal_2007A&A...472..835Z} located at $\alpha_{J200}$=17$^{h}$12$^{m}$20.6$^{s}$, $\delta_{J200}$=-38$^{\circ}$29'26''.  This star is found in a Galactic HII region  embedded in a molecular cloud  with numerical density of $n_a \sim$ 3000 cm$^{-3}$ \citep{Zavagno_etal_2007A&A...472..835Z}. {The age of the} massive
star is $<$ 3 Myr, and the dynamical age of the shell is $\sim$0.2 Myr \citep{Arthur2011MNRAS.414.1747A}. More recently, \citet{Torii_etal_2015ApJ...806....7T} found that the dynamical age is $\sim$0.4 Myr. \citet{Ochsendorf_Cox_2014A&A...563A..65O} proposed a pressure and density gradient in the ISM around the ionising source of RCW\,120, in order to explain the asymmetry of the HII region and the MIR arcs. However the diameter of the RCW~120 is around 3.8 pc \citep{Anderson_etal_2015ApJ...800..101A} and a 
pressure or density gradient must be related to a stellar motion \citep{Weaver_etal_1977ApJ...218..377W, 1986ApJ...300..745R, 1991ApJ...369..395MacLow, Arthur_Hoare_2006ApJS..165..283A, 2013MNRAS.436..859M, Steggles_etal_2017MNRAS.466.4573S}. Most probably, the stars, particularly the massive stars, are born
in motion with respect to their surroundings because they keep the momentum gained during their star formation process, where the turbulence is needed, with velocities of 2$-$5 km s$^{-1}$ \citep[e.g.,][] {2010ApJ...719..831P, Dale_Bonnell_2011MNRAS.414..321D}. 

Many authors have modelled bow shocks from runaway O
and B stars \citep[e.g. ][]{1991ApJ...369..395MacLow, Comeron_Kaper_1998A&A...338..273C,2014MNRAS.444.2754Meyer} and there have been a few studies of H II regions
for $v \geq$20 km s$^{-1}$ , \citep{1979A&A....71...59T, 1997RMxAA..33...73Raga, 2013MNRAS.436..859M}. In both cases a partial shocked shell
forms, upstream for the bow shock and in the lateral direction for
the H II region, and can be unstable in certain circumstances. {In the case of RCW 120, \citet{2016A&A...586A.114M} performed 2D radiation
hydrodynamical numerical simulations of stellar wind bubbles and
post-processing calculations to obtain synthetic intensity maps at infrared
wavelengths. They found good agreement between their model and the RCW 120
observations. In particular, they obtained an infrared arc
with the same shape and size as the arc observed in RCW 120, with a
brightness comparable to the observed one. They concluded that their results
could be showing the outer edges of an asymmetric stellar wind bubble.}
%
{On the other hand, the respective effects of the smaller and  larger dust grains has been studied by \citet{vanMarleetal2011ApJ...734L..26V}. These latter authors modelled the gas and dust dynamics using high-resolution
numerical simulations of a fast-moving red supergiant and concluded 
that the smaller sized grains follow the motion of the gas and the larger grains are weakly
coupled to the gas.
Therefore, they proposed that the small dust grains dominate the emission in the infrared observations. Moreover, \citet{2006A&A...449.1061D}  and \citet{2017MNRAS.464.3229M}  showed that a small magnetic  field can
affect the predicted emission from the post-shock recombination regions by one or two orders of
magnitude. However, the work of these latter authors was focused on the
optical emission prescriptions and they did not 
study the infrared emission mainly produced by the dust grains.  Indeed, \citet{2017MNRAS.464.3229M} made
post-processing calculations of the dust continuum infrared emission and
showed that the presence of the magnetic field decreases the infrared emission.
Finally, using 3D magneto-hydrodynamical  
simulations \citet{2018MNRAS.473.1576K} showed that the
drag force cannot be neglected in 
models with ISM densities larger than 10 cm$^{-3}$.}\\
{More recently, \cite{2019MNRAS.486.3423H,2019MNRAS.486.4423H,2019arXiv190400343H}, studied the formation of IR-emission
arcs around luminous stars that move supersonically
relative to their surrounding medium. These latter authors found four different regimes depending on the type of coupling
between dust grains and gas, and taking into consideration the relative
values between the stellar wind and the ultraviolet optical depth of the
bow shell. They presented cases in which a radiation-supported dust wave decouples from the gas to form an
infrared emission arc outside of any hydrodynamic bow shock, considering the
magnetic fields in their numerical models.
They considered the drag force because of the dust and the gas interactions assuming that the fraction of the dust is zero in the wind.}\\
%
In the present study, we modelled the RCW~120 physical properties, performing a series of numerical simulations that solve the radiation-hydrodynamics and dust equations in a fully 3D fixed cartesian mesh. 
This paper is organized as follows: the numerical simulation setup is described in
Sect. 2, the simulation results are presented in Sect. 3, and finally we conclude in Sect. 4.
%
\section{Governing equations}

We performed 3D simulations using the Guacho\footnote{The Guacho code is a free access code which
is maintained in http://github.com/esquivas/guacho} code \citep{Esquivel_etal_2009A&A...507..855E,Esquivel_etal_2013ApJ...779..111E}. The Guacho code uses a Cartesian grid to solve the near-conservation laws governing the gas-dynamics and the neutral hydrogen rate.  
It includes the radiative transfer at the Lyman limit of ionising photons emitted from an isotropic point source photoionising a neutral hydrogen distribution. Also, it uses the ray-tracing scheme where the
diffuse radiation is considered only in the B recombination case. The code is described in detail in \citet{Esquivel_etal_2013ApJ...779..111E} and in \citet{Schneiter_etal_2016MNRAS.457.1666S}. In this project we incorporate the dust-grain dynamics coupled to gas dynamics and the momentum feedback of radiation on the dust.

The radiative cooling is included as described by \cite{Raga_Reipurth2004RMxAA..40...15R} for the atomic gas, and for lower temperatures we have included the parametric  molecular cooling function presented by \citet{Kosinski_Hanasz2007MNRAS.376..861K} and  \citet{Rivera-Ortiz_etal_2019ApJ...874...38R}. The complete set of equations is as follows.

\begin{eqnarray}
 \frac{\partial \rho}{\partial t}+\vec{\nabla} \cdot \left(\rho\,\vec{v}\right)&=&0,\label{eq:gas:density}\\
 \frac{\partial \rho v}{\partial t}+\vec{\nabla} \cdot \left(\rho \,\vec{v}\cdot \vec{v}+p\right)&=&\rho\,f_{\rm rad,\, nHI}+f_{\rm d},\\
\frac{\partial e }{\partial t} +\vec{\nabla}\cdot \left[\vec{v} \left(e+p\right) \right] &=&G_{\rm rad}-L_{\rm rad}+f_{\rm rad,nHI}\;\vec{v}
\,+\,f_{\rm d}\,\vec{v}\,,\label{eq:ener}\\
\frac{\partial \rho_{\rm d}}{\partial t}+\nabla \left(\rho_{\rm d}\vec{v_{\rm d}}\right)&=&0,\label{eq:dust:density}\\
\frac{\partial \rho_{\rm d} v_{\rm d}}{\partial t}+ \nabla(\rho_{\rm d} \vec{v_{\rm d}}\cdot \vec{v_{\rm d}})&=&\rho_{\rm d} f_{\rm rad,d}\,-\,f_{\rm d}\label{eq:dust:momentum},
\end{eqnarray}

%
\noindent where, $\rho$, $\vec{v}$, $p,$ and $E$ are the density, the velocity, the thermal pressure and the total energy density of the gas, respectively.  $G_{\rm rad}$ and $L_{\rm rad}$ are the
energy gains and losses due to radiation. We assume a standard ideal gas law for closure, and therefore 
the gas total  energy density and
the gas thermal pressure are related by $E=\rho \vec{v}^2/2+P/(\gamma-1)$, where the ratio between specific capacities is $\gamma=5/3$. 

The factors $\rho_{\rm d}$ and $\vec{v}_{\rm d}$ are the density and the velocity of the dust. The drag force $f_{\rm d}$ is given by a combination of Epstein's drag law for the subsonic regime and Stokes' drag law for the supersonic regime  \citep[see,][]{vanMarleetal2011ApJ...734L..26V},

\begin{equation}\label{eq:f_drag}
    f_{\rm d} \,=\,0.57\,\pi\,\frac{\rho_{\rm d}}{m_{\rm d}}\rho\,a_{\rm d}^2\left(v-v_{\rm d}\right)\sqrt{\left(v-v_{\rm d}\right)^2+c_{\rm s}^2},
\end{equation}
where $c_{\rm s}=\sqrt{\gamma\frac{p}{\rho}}$ is the gas sound speed. {The test of the hydrodynamics equation coupled with dust using the drag forces in Eq. \ref{eq:f_drag} shows the same results as those obtained by \citet{vanMarleetal2011ApJ...734L..26V}.}
Finally, the radiation pressure gradient force   due to photoionisation is
\begin{equation}
f_{\rm rad,nHI}\,=\,n_{\rm HI} \phi_{\rm H} \frac{\chi_H}{c},
\end{equation}
and  the radiation pressure gradient force due to grain dust  absorption is
\begin{equation}
f_{\rm rad,d}\,=\,\frac{n_{\rm d} \sigma_{\rm d} F^*}{c} e^{-\Delta \tau_{\rm d}} ,
\end{equation}
where $\phi_{\rm H}$ is the Lyman $\alpha$ photoioinzation rate, $\chi_{\rm H}$ is the
hydrogen ionization potential, $\sigma_{\rm d}$
is the grain dust cross-section (the geometrical cross section),
$F^*$ is the photons flux, and 
$c$ is the light speed. More details about the radiative force are given in \citep{Rodriguez-Ramirez_2016ApJ...833..256R}.

In the ray tracing scheme, the photon  package emitted by an isotropic point source 
 decays by a factor of e$^{-\left(\Delta \tau_{\rm H} + \Delta \tau_{\rm d}\right)}$ when travelling through neutral gas and dust. Along a path-length $\Delta l$,  the opacities induced by neutral gas and dust are $\Delta \tau_{\rm H}\,=\,{\rm a}_0\,{\rm n}_{\rm HI}\, \Delta l$ and $\Delta \tau_{\rm d}\,=\,{\rm a}_{\rm d}\,{\rm n}_{\rm d}\Delta l$, respectively. Here, ${\rm a}_{0}$ is the  photoionisation cross-section constant, ${\rm a}_{\rm d}$ is the geometrical dust cross-section, and ${\rm n}_{\rm HI}$ and ${\rm n}_{\rm d}$ are the HI and dust number density. The photoionisation rate $\phi_{\rm HI}$ is  computed by equating the ionising photon rate, $S$, to the ionization per unit time in each cell, that is, $S\,=\,n_{\rm HI} \phi_{\rm HI} dV$. 

\subsection{Simulation parameters and setup}
The coupled gas--dust and radiation pressure equations are advanced using the HLL scheme from \citet{Hartenetal83}  with the second-order spatial re-construction minmod slope limiter coupled with the second-order time step.

The presented hydrodynamical simulations (Table.\ref{tab:models}) are carried in a Cartesian grid with a physical domain $\left[x,y,z\right]\, \in\,\left[\left(0,3\right),\left(0,3\right),\left(0,9\right)\right] {\rm pc}$ with a spatial resolution of $0.03\, {\rm pc}$ in each direction.



The ISM has a uniform number density of ${\rm n}_{\rm a}=3000$~cm$^{-3}$, temperature ${\rm T}_0\,=\,100\,{\rm K,}$ and is moving  in the {\it z} direction with 
a velocity of 5~km~s$^{-1}$ (to emulate the stellar motion of RCW\,120).
Our numerical models considered appropriate
values of the stellar wind velocity (2300~km~s$^{-1}$) and a mass injection rate (2.7$\times 10^{-7}$~M$_\odot$~yr$^{-1}$ for one single O8V star  \citep{Sternberg_2003ApJ...599.1333S}. The steady spherical stellar wind was imposed in a sphere of radius $0.09\,{\rm pc}$ at the simulation grid centre. 

We used a  photon ionisation rate of  1.6$\times$10$^{48}$~phot~s$^{-1}$ (see \citealt{Zavagno_etal_2007A&A...472..835Z}) and the photons are injected in 
the surface of a sphere with a radius of 1.5 times the wind source radius.

We assume the dust to be composed of silicates, for which the internal particle density is  $\rho_{\rm grain}=3.3\,{\rm g\, cm^{-3}}$ \citep{Draine_Lee1984ApJ...285...89D}. Concerning the dust shape, we assume spherical grains with a radius of $a_{\rm dust}$. The simulations are performed with two dust radii, $a_{\rm dust}\,=\,0.1$ and $a_{\rm dust}\,=\,1~\mu {\rm m}$.  The dust to gas fraction  is $\chi=\frac{\rho_{\rm d}}{\rho_0}\cdot\rho_{\rm grain}*V_{\rm grain}$, where
$V_{grain}$ is the volume of the grains. The dust is only included into the uniform ISM. 

We computed simulations of RCW\,120, varying the dust fraction and size,
the momentum of the wind, the number of ionising photons, and the density in
the ISM (see Table~\ref{tab:models}). Following \citet{2018RMxAA..54..375S} and \citet{Torii_etal_2015ApJ...806....7T}, 
we ran our simulations up to a time of 0.4 Myr.

\begin{table}
\label{tab:doi}
\begin{center}
\caption{Characteristics and parameters of the simulations. We show the dust fraction $\xi$, the dust radius a$_d$ [cm], the mass-loss rate for the model M5 (which is different from the others) and the absence of photons in the model M1. In the cases M1-M4 the mass flux is  2.7$\times 10^{-7}$~M$_\odot$~yr$^{-1}$ , and in the M2-M5 cases the ionising photon rate is 1.6$\times$10$^{48}$~phot~s$^{-1}$. In the model M2 there is no dust.}
\begin{tabular}{c  c c l}
\hline \hline
Model       &
\multicolumn{2}{c}{Dust1}  & comments\\
          & $\chi_{\rm 1,ISM}$   &   a$_d$ [cm] &\\ 
\hline
M1      & $0.02$  & 10$^{-4}$ & no photons  \\ 
M2       & $0.00$  & $---$  &    \\
M3        & $0.02$  & 10$^{-4}$ &    \\
M4      & $0.02$  & 10$^{-5}$  &  \\ 
M5      &  $0.02$  & 10$^{-4}$ &    $\dot{\rm M}=10^{-2} \dot{\rm M}_w$\\ 
\hline\hline
\end{tabular}
\label{tab:models}
\end{center}
\end{table}

\section{The gas and dust decoupling}

\begin{figure}
\begin{center}
\includegraphics[width=0.7\columnwidth]{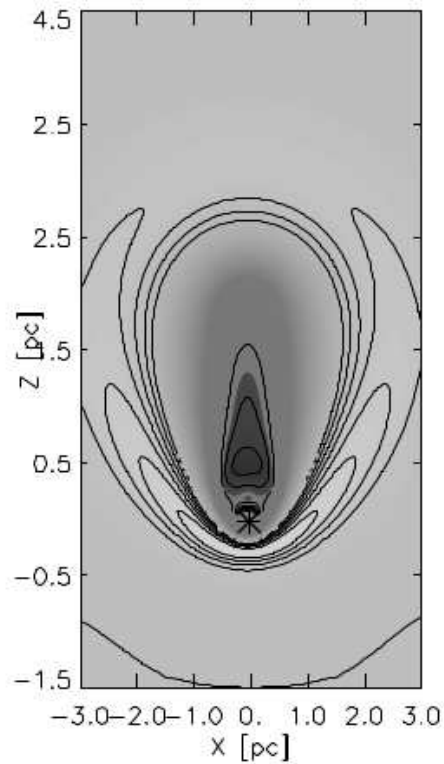}
\caption{Two-dimensional cut in  the Z-X plane of the  density contour in grey and dust contour with continuous lines.}
\label{fig:xzcut}
\end{center}
\end{figure}

To understand the stellar wind and the photon pressure effect in the gas and dust dynamics and the dust--gas interaction, we performed five numerical models  (Table~\ref{tab:models}). A cut in the $Z-X$ plane of the model M3 is shown in Figure~\ref{fig:xzcut}. As we can see, two shock regions are formed: one strong shock in front of the runaway star, the nearest one; and the second, a weaker shock behind the runaway star. In  the tail  of the HII region, the cometary HII region, the shock wave forms a shell of less-dense, shocked ISM.

\begin{figure}
\includegraphics[width=\columnwidth]{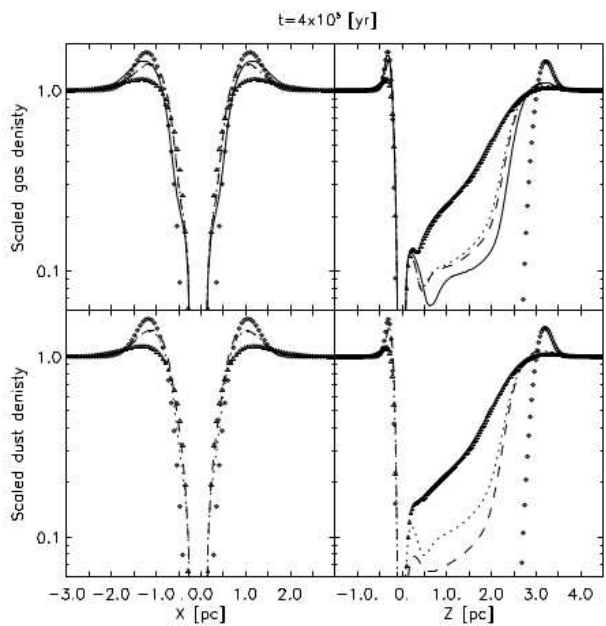}
\caption{Scaled gas density (upper panels) and scaled dust density (lower panels) along the $X$-axis (perpendicular to the  direction of star propagation) and $Z$-axis (along the moving direction), at 0.4 Myr. The diamonds correspond to the M1 model, the solid, dotted, and dashed lines correspond to
models M2, M3, and M4, and the triangles
are for the M5 model, respectively.}
\label{fig:zmodels}
\end{figure}

\begin{figure}
\includegraphics[width=\columnwidth]{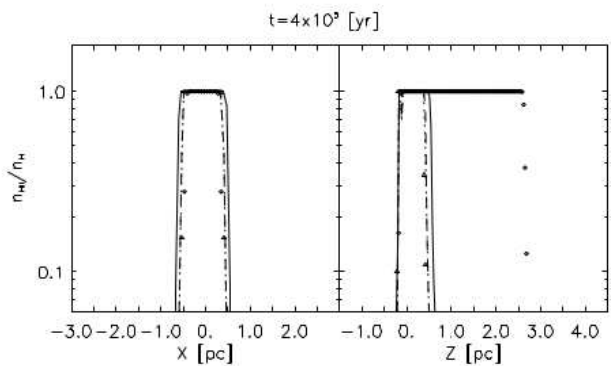}
\caption{Ionisation fraction for the M1, M2, M3, and M4 models at 0.4~Myr for a cut on the $X$ and $Z$ axes (left and right panel, respectively) in the centre of the simulation box. Lines and symbols are as in Figure~\ref{fig:zmodels}.}
\label{fig:yzfrachii}
\end{figure}

%
In order to analyse the behaviour of these two regions, we show in Figure~\ref{fig:zmodels} a 1D cut along two directions: a cut perpendicular to the moving star direction, the $X$-axis, and a cut in the moving direction, the $Z$-axis; at time $0.4$ Myr (left an right panels of Figure~\ref{fig:zmodels} respectively). It is clear that the two regions mentioned above are observed.
For the M1 model, where only the stellar wind is considered without including the photons flux (the diamonds line in Figure~\ref{fig:zmodels}), the gas and dust are well coupled and  are stacked at the external shell.
Inside the  bubble within the shell of the shocked ISM, there is a rarefied region with low gas and dust density. However, the gas is hot in this region. In Figure~\ref{fig:yzfrachii} we show the fraction of ionised gas for cuts on the $X$ and $Z$ axes (left and right panels, respectively). In the cut on the $X$ axis, one can see that the gas is completely ionised within a narrow region where the ram pressure and the over-pressure of the hot bubble reach equilibrium. This bubble of ionised (hot) gas is much larger in the $Z$ direction  than in the $X$-axis (1.2~pc.), namely about 3~pc extending in an opposite direction to the movement of the star, giving the cometary shape in this region. For this model, the ionised region has an ellipsoidal (egg-like) shape.
In  the other models, M2-M5,  the ionising photons are included (Table~\ref{tab:models}). In those models, the ISM is heated by the ionization front which decreases the shock Mach number resulting in a lower compression in the external shell. At the tail of the cometary HII region only a compression wave forms and the shock remains only at the front and the side of the runaway star (Figure~\ref{fig:zmodels}). As a result, at the back of the moving star inside the bubble, the density  increases  and the temperature decreases.
%
Models M3-M4, with dust in ISM and photoionisation, show a greater difference behind the runaway star inside the bubble, where more ISM gas and dust remains. Indeed, the dust decreases the compression wave strength. Moreover, the structure of this  rarefied zone changes under the effect of the dust (Figure~\ref{fig:zmodels}). There is an increment in the gas density to 0.25~pc from the star and a minimum (local) at 1~pc for M3 and M4 models in comparison with 1.1~pc for the M2 model. 
%
Concerning the dust distribution in the M1, M3, M4, and M5 models, in front of the runaway star (Figure~\ref{fig:zmodels}) the compression shock is strong and the dust is tightly coupled with the gas. However, inside the bubble at the back of the runaway star in the rarefied region the dust is decoupled from the gas. This decoupling between the dust and gas is more important in model M3 because the dust grains are bigger. 
%
In models M3 and M4 there are zones where the dust accumulates. The denser zone is at the forward shock and the second is inside the bubble in the rarefied region (Figure~\ref{fig:zmodels}). As a result, there are two regions with a significant dust grain emission: a) an outer  region of swept up ISM and b) a region more  internal and close to the ionising star, at approximately 0.2~pc, which is formed with ISM gas filling the decreased pressure region due to the star motion.  
%

\section{Dust distribution in RCW~120}
\begin{figure}
\includegraphics[width=\columnwidth]{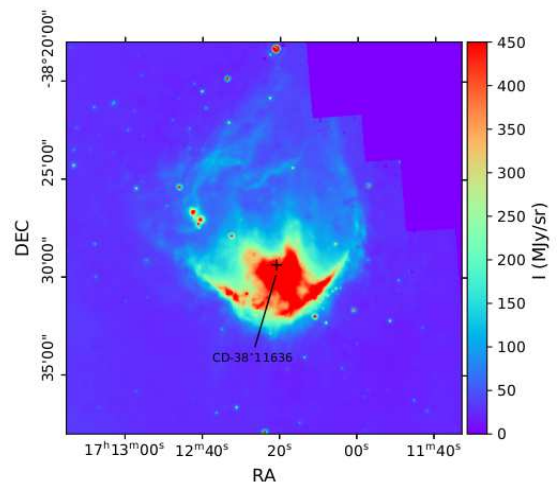}
  \caption{RCW~120 image at 24 $\mu$m from Spitzer-MIPSGAL \citep{Carey_etal_2009PASP..121...76C}, showing emission of small dust grains heated by the stellar radiation. }
  \label{fig:ir}
\end{figure}

\begin{figure}
\includegraphics[width=\columnwidth]{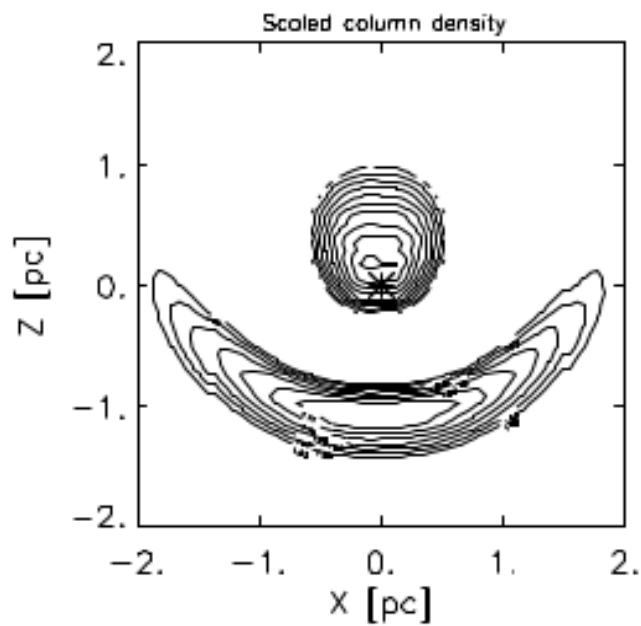}
  \caption{Projection plane X-Z image of the scaled column density. Contours represent the scale column density for dust. The contours around the star have values around 0.1, while in the shell the rank is between 1.65 and 2.0}
  \label{fig:map_rot}
\end{figure}

A visual comparison of our models is described in this section. Figure \ref{fig:ir} shows the emission of the Galactic RCW~120 bubble at  24 $\mu$m from Spitzer-MIPSGAL \citet{Carey_etal_2009PASP..121...76C}. MIPSGAL is a survey of the  same region as  Spitzer-GLIMPSE surveys (see \citealt{Churchwell_etal_2009PASP..121..213C} for more details) using the  Multi-band Imaging Photometer for Spitzer \citep[MIPS;][]{Rieke_etal_2004ApJS..154...25R} instrument  at 24 $\mu$m. The  MIPSGAL  resolution is  6'' at  24 $\mu$m. This image shows the extended emission of small dust grains at 24 $\mu$m in the region of the ionised gas and the emission in the envelope associated with young stellar objects (YSOs).
As one can see, the emission at 24 $\mu$m of dust is mainly localised in two places: a) in an external and thin shell in the front part of the cometary HII region, and b) near the ionising source in a central region of the bubble that has a significant dust fraction. Figure~\ref{fig:xzcut}
shows a cut on the X-Y axis for the M3 model; the grey map is the gas density and the dust density contours. The blue contours correspond to the normalized density of dust larger or equal to 0.8 (the outer shell) and 
the contour corresponds to densities lower than 0.8 (the inner dust
structure).
As we can see in the Figure~\ref{fig:xzcut}, the HII region has an ellipsoidal  (egg-like) structure and does not look like a spherical shape as shown in the observation in Figure~\ref{fig:ir}. In addition, although the distribution of the dust is in the external shell, because it was dragged by the leading shock, it is also seen that the internal structure of dust is behind the position of the source (marked with an asterisk in Figure~\ref{fig:xzcut}).
\citet{Marsh_Whitworth_2019MNRAS.483..352M} determined the shape of the external shell  of RCW~120 and  concluded
that it is a circular, and not elongated, ring with an internal radius of 1.66~pc and a
depth of 0.33~pc. The elongation presented in the current numerical models with a star motion cannot explain the shape of the RCW~120. \citet{Torii_etal_2015ApJ...806....7T} studied the 
proper motion in RCW~120 and found a problem with the determination of the molecular shell velocity with respect to the observer.  It is therefore likely that the object has a projection angle that does not allow us to see an elongated shape
in this HII region.
Finally, in order to compare our model with observations, we built column density maps, $N(i,j)_ {columnar} = \sum n(i, j) * dR$, where {a rotation has been included}. Figure~\ref{fig:map_rot}
shows the column density for the dust for a rotating angle of 72  degrees of the projection plane X-Z.  
If the line of vision is perpendicular to the axis of movement, the external shell has the shape of a ring, with the ionising 
star at the centre of it. However, the rotation of 72 degrees that we have used results in an external shell with a 
spherical section in the direction of the movement of the star. Also, the internal region with dust now coincides with 
the position of the star.

\section{Conclusions}
We present several numerical models for the evolution of the HII region, RCW~120. Our models consider the photoionisation effect, the wind injected by the central star, and the effect of dust on the dynamical evolution of
this object and the stellar motion.

With respect to the M1 model, (without photoionisation), the gas and dust do not seem to uncouple, since they are dragged by the leading shock and both of them are stacked in an external shell.
We performed a numerical simulation for a weak wind model (where the stellar wind transfers a momentum reduced by a factor of 100 compared with the other models). This model, M5, produces a bubble with gas and dust and  
a significant decoupling of the dust and gas is not seen.

Our strong stellar winds models show an internal region with a high density of gas and dust very close to the central star. On the other hand, the distribution of dust and gas inside the bubble is different, with a partial decoupling between them. The model in which the dust grains are  smaller maintains the coupling or partially  coupling of gas and dust for a longer time despite the interactions with  the ionisation and shock fronts. The decoupling is greater in models M3 and M4, with larger dust grains 1~$\mu$m. In these  models, the dust is mainly concentrated in the external shell and in the central structure near the position of the star. That is, the gas and dust density profiles are found with a very similar distribution as inside the RCW~120 bubble.

We show that when making a rotation of 72 degrees, the morphology results in an external shell with a spherical ring shape. Surprisingly, the internal region with dust now coincides with the position of the star. This coincides with the observations which suggest that there is an angle with respect to the line of sight that presents us from observing an elongated shape in RCW~120.

As far as we are aware, this is the first time that {3D} gas-dynamic equations, coupled with the photoionisation and the dust together with the stellar motion, have been solved for RCW~120. Therefore, according to our models, {all of} these  are of great importance to understand the distribution of dust in the RCW~120 bubble.

{ Our results show good agreement with those obtained by \citet{2016A&A...586A.114M},
and are in accordance with the general analysis of \citet{2019MNRAS.486.4423H} for the case
of a dust wave where the dust is decoupled from the gas. Indeed, the
distribution of dust is mainly located in a region found immediately after the bow
shock, as in the two works mentioned previously.}

{We highlight the fact that we have not included the magnetic field in our simulations coupled with radiation
hydrodynamics and dust, and that this is important in order to compute synthetic maps with optical
and infrared emission. The inclusion of the magnetic field in simulations and the effect of this on the emission is left for future work}.

Finally, we have shown the importance of strong stellar wind in the decoupling of the gas and dust during the lifetime of the  bubble.

\section{Acknowledgements}
We acknowledge support from DGAPA-UNAM grants IN-109518 and IG-100218. 
ZM is grateful for the use of the High Performance Computing OCCIGEN  
CINES within the DARI project A0030406842.

\bibliographystyle{aa}
\bibliography{references}

\end{document}